\documentclass[reprint,superscriptaddress,amsmath,amssymb,aps,prl, longbibliography,floatfix]{revtex4-2}
\usepackage{float}
\usepackage{graphicx}
\usepackage[svgnames]{xcolor}
\usepackage[bookmarks=false]{hyperref}
\usepackage{dcolumn}
\usepackage{bm}

\usepackage{natbib}
\begin{document}
\title{Formation of an electron-phonon bi-fluid in bulk antimony}

\date{\today}
\author{Alexandre Jaoui}
\email{alexandre.jaoui@espci.fr}
\altaffiliation{\\Present address : ICFO - Institut de Ciencies Fotoniques, The Barcelona Institute of Science and Technology, Castelldefels, Barcelona, Spain.}
\affiliation{JEIP, USR 3573 CNRS, Coll\`ege de France, PSL Research University, 11, Place Marcelin Berthelot, 75231 Paris Cedex 05, France.}
\affiliation{Laboratoire de Physique et d'\'Etude des Mat\'eriaux \\ (ESPCI - CNRS - Sorbonne Universit\'e), PSL Research University, 75005 Paris, France.}

\author{Adrien Gourgout}
\affiliation{Laboratoire de Physique et d'\'Etude des Mat\'eriaux \\ (ESPCI - CNRS - Sorbonne Universit\'e), PSL Research University, 75005 Paris, France.}

\author{Gabriel Seyfarth}
\affiliation{LNCMI-EMFL, CNRS, Univ. Grenoble Alpes, INSA-T, UPS, Grenoble, France.}

\author{Alaska Subedi}
\affiliation{CPHT, CNRS, Ecole Polytechnique, IP Paris, F-91128 Palaiseau, France}
\affiliation{Coll\`ege de France, 11 Place Marcelin Berthelot, 75005 Paris, France}

\author{Thomas Lorenz}
\affiliation{II. Physikalisches Institut, Universit\"at zu K\"oln, 50937 K\"oln, Germany}

\author{Benoît Fauqué}
\affiliation{JEIP, USR 3573 CNRS, Coll\`ege de France, PSL Research University, 11, Place Marcelin Berthelot, 75231 Paris Cedex 05, France.}

\author{Kamran Behnia}
\email{kamran.behnia@espci.fr}
\affiliation{Laboratoire de Physique et d'\'Etude des Mat\'eriaux \\ (ESPCI - CNRS - Sorbonne Universit\'e), PSL Research University, 75005 Paris, France.}

\begin{abstract}
The flow of charge and entropy in solids usually depends on collisions decaying quasiparticle momentum. Hydrodynamic corrections can emerge, however, if most collisions among quasiparticles conserve momentum and the mean-free-path approaches the sample dimensions. Here, through a study of electrical and thermal transport in antimony (Sb) crystals of various sizes, we document the emergence of a two-component fluid of electrons and phonons. Lattice thermal conductivity is dominated by electron scattering down to 0.1 K and displays prominent quantum oscillations. The Dingle mobility does not vary despite an order-of-magnitude change in transport mobility. The Bloch-Gr\"uneisen behavior of electrical resistivity is suddenly aborted below 15 K and replaced by a quadratic temperature dependence. At Kelvin temperature range, the phonon scattering time and the electron-electron scattering time display a similar amplitude and temperature dependence. Taken together, the results draw a consistent picture of a bi-fluid where frequent momentum-conserving collisions between electrons and phonons dominate the transport properties.
\end{abstract}

\maketitle

\section{Introduction}

Hydrodynamic corrections~\cite{gurzhi1968} to transport properties of solids can emerge when the travelling quasiparticle endures momentum-conserving collisions outweighing the momentum-relaxing ones. Their signatures have been reported for electrons in mesoscopic metals~\cite{molenkamp1994,moll2016,crossno2016,bandurin2016,gooth2018,sulpizio2019} and for phonons in bulk insulators~\cite{beck1974,martelli2018,machida2018,machida2020}. The possible role played by phonons in the emergence of hydrodynamic effects in the electronic fluid~\cite{hartnoll2007} has become a subject of recent theoretical attention~\cite{Coulter2019,vool2020,levchenko2020,lucas2021}. It remains to be seen if momentum exchange between electron and phonon baths can generate experimental signatures other than phonon drag~\cite{fu2018}, the well-known non-diffusive thermoelectric response of heat-carrying phonons coupled to the electron bath~\cite{Herring1954}.  

\begin{figure*}
\begin{center}
\makebox{\includegraphics[width=1\textwidth]{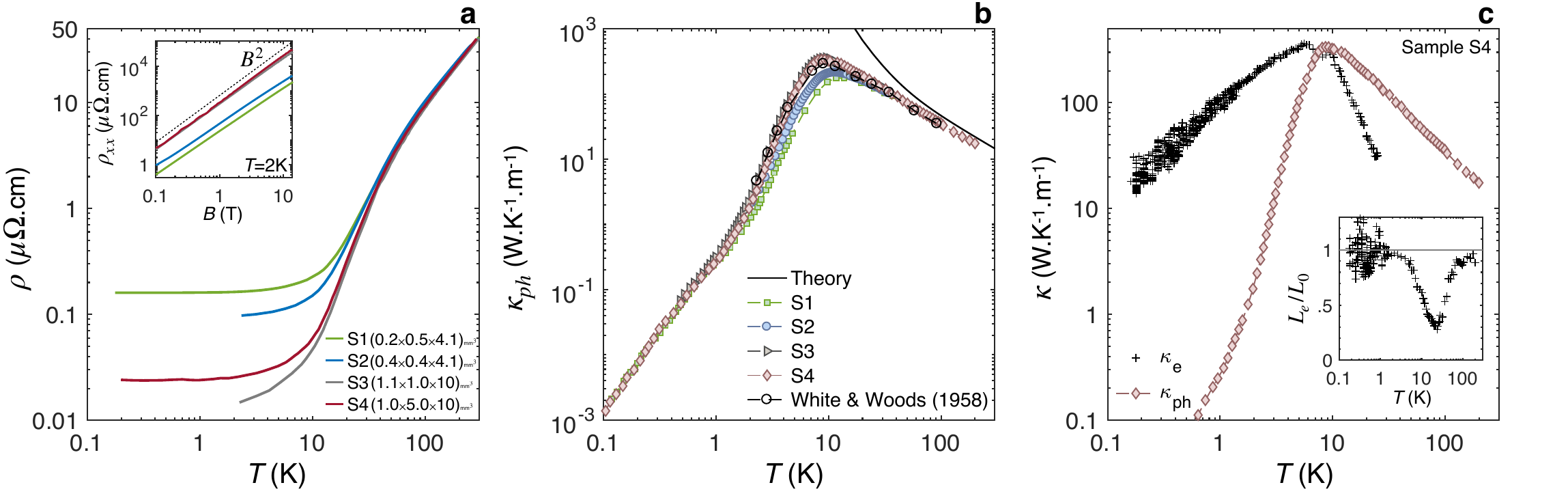}}
    \caption{\textbf{Electrical and thermal conductivity in Sb crystals along the bisectrix crystallographic orientation. a} Temperature dependence of the electrical resistivity $\rho$ of four different Sb crystals of different sizes. Larger samples show lower residual resistivity. The inset shows the magnetoresistance of these crystals at $T$=2 K. \textbf{b} Lattice thermal conductivity of the same four Sb samples. Data from a previous report on a Sb rod \cite{white1958} is also included. The solid line represents theoretical lattice thermal conductivity considering only three-phonon scattering events (and neglecting four-phonon scattering as well as scattering by electrons). Note that the opening gap between experiment and theory with cooling. \textbf{c} Temperature dependence of the electronic and phononic contributions to the total thermal conductivity in sample S4. The procedure for separation is discussed in detail in the supplement \cite{SM}. The inset shows the temperature dependence of the electronic Lorenz number $L_e=\kappa_e\rho/T$ normalized by $L_0$, the Sommerfeld value.}
    \label{fig1}
\end{center}
\end{figure*}

Bulk semimetals are promising platforms for this investigation. Their small Fermi surface pockets implies reduced Umklapp electron-electron scattering. Transport studies have found that a significant fraction of electron-electron scattering conserve momentum in WP$_2$ ~\cite{gooth2018,jaoui2018} and in Sb~\cite{jaoui2021}, generating a downward departure from the  Wiedemann-Franz (WF) law around \textit{T}$\sim$10 K, near the onset of the ballistic regime. Magnetic imaging experiments~\cite{vool2020} find a Poiseuille profile of electron flow in the same temperature range in WTe$_2$. According to theoretical calculations, phonons play a prominent role in the emergence of this hydrodynamic window~\cite{vool2020}, a conjecture supported by Raman scattering experiments in WP$_2$~\cite{Osterhoudt2021}.

Here, we present direct evidence for an unprecedented case of phonon-electron coupling in bulk antimony. We find that the thermal diffusivity of phonons displays a non-monotonous temperature dependence with a Poiseuille peak as observed in several other solids. In contrast to all other cases, however, instead of entering a ballistic regime upon further cooling, the phonons continue to exchange momentum with charge carriers down to 0.1 K and display quantum oscillations of their thermal conductivity. The phononic viewpoint from thermal transport is complemented by the electronic viewpoint studied from electrical transport. As expected in the Bloch-Gr\"uneisen picture of electron-phonon (e-ph) scattering, the exponent of resistivity increases when the system is cooled below the Debye temperature, but the steady enhancement is suddenly interrupted. Below \textit{T}$\approx$15 K, the electrical resistivity becomes purely $T$-square. We argue that this is because the smallness of phonon wavevector does not allow Umklapp events and therefore e-ph scattering does not contribute to resistivity. Around 1K, the amplitude and the temperature dependence of the phonon scattering time (extracted from phonon conductivity) and the e-e scattering (extracted from electronic transport) match each other in  amplitude and temperature dependence. The most plausible explanation for this is that colliding electrons exchange phonons as previously suspected~\cite{vool2020}. 

Our scenario is backed by  comparisons between the transport properties of antimony and metals with larger (i.e. copper or tungsten) and lower (bismuth or black phosphorous) carrier densities. There is roughly one charge carrier per 1000 atoms in antimony. The Fermi surface is small enough to push the mean-free-path of electrons close to ballistic, but also sufficiently large to render electrons capable of scattering phonons when phonon-phonon collisions cease to decay the heat current well below the Debye temperature. 

\section{Results}
Fig.\ref{fig1} shows the temperature dependence of electrical resistivity and thermal conductivity in antimony. As discussed in the supplement~\cite{SM}, the electronic $\kappa_e$, and the phononic $\kappa_{ph}$ components of the total thermal conductivity, $\kappa$, can be easily separated thanks to the large sensitivity of $\kappa_e$ to magnetic field (a consequence of the huge magnetoresistance) and the field independence of $\kappa_{ph}$. As seen in Fig.\ref{fig1}.b, $\kappa_{ph}$ peaks at $T \sim 10$ K and (in contrast to the strong size dependence of $\kappa_{e}$~\cite{jaoui2021}) does not show any dependence on the sample size at low temperature. We will discuss the implications of this feature below. This figure also shows the theoretical $\kappa_{ph}$ computed from the phonon spectrum. The calculations have neglected scattering by electrons, the finite sample size and defects.  They are also restricted to three-phonon scattering events. As seen in the figure, the experimental and the theoretical $\kappa_{ph}$ are close to each other at room temperature. However, there is a deficit in the measured $\kappa_{ph}$, compared to the predicted one. The difference grows with cooling.  Similar theoretical calculations yield a quantitaive account of the experimental data down to low temperature in Si \cite{broido07}, GaAs \cite{lindsay13}, PbTe \cite{shiga12}, SnSe \cite{guo15}, Al$_2$O$_3$ \cite{dongre18}, and In$_2$O$_3$ \cite{subedi21}.  Scattering of phonons by electrons is the most plausible reason that this is not the case of antimony.

\begin{figure*}
\begin{center}
\makebox{\includegraphics[width=1\textwidth]{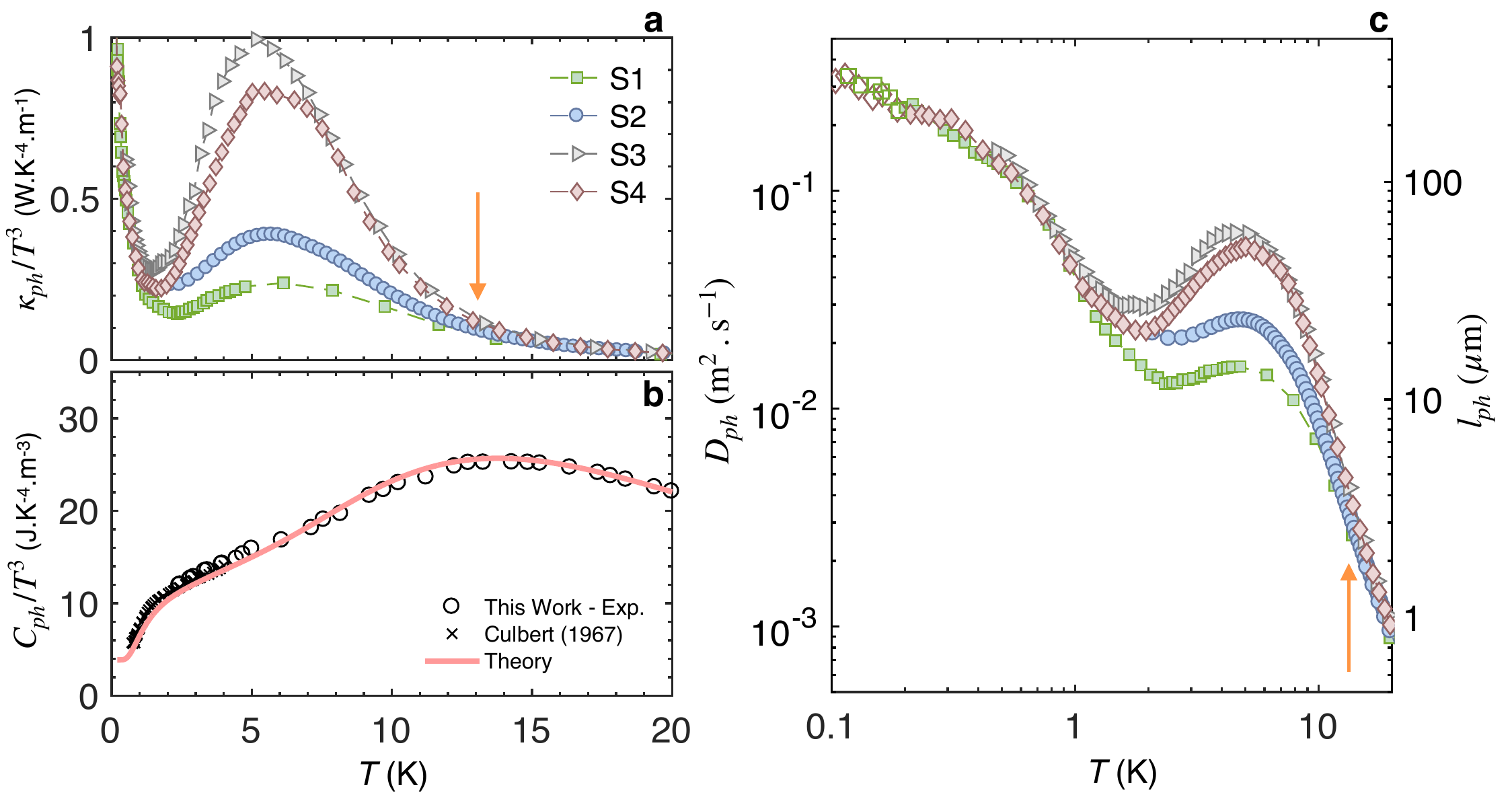}}
    \caption{\textbf{Thermal conductivity, heat capacity and thermal diffusivity of phonons. a} Phononic thermal conductivity $\kappa_{ph}$, plotted as $\kappa_{ph}/T^3$ as a function of temperature. \textbf{b} The phononic specific heat $C_{ph}$ divided by $T^3$ together with a previous report~\cite{culbert1967}. The solid line represents \textit{ab initio} calculations. \textbf{c} The lattice thermal diffusivity $D_{ph}$ (left \textit{y}-axis) and phonon mean-free-path $l_{ph}$ (right \textit{y}-axis) \textit{vs.} temperature. Note the emergence of a size-dependent  peak and a  minimum in the $T$=2-15 K temperature range. Orange arrows indicate the temperature below which $D_{ph}$ and  $\kappa_{ph}$ become size-dependent.}
    \label{fig2}
\end{center}
\end{figure*}

 In the vicinity of its peak, $\kappa_{ph}$ depends on the sample size. This is more clearly seen in Fig.\ref{fig2}.a, which shows the temperature dependence of $\kappa_{ph}/T^3$ and its prominent peak. Fig.\ref{fig2}.b), shows the lattice specific heat $C_{ph}$ below 20 K. It does not follow a $T^3$ behavior, but is in excellent agreement with our first-principle calculations of the phonon spectrum  (see the details in the supplement\cite{SM}). This agreement implies  that the discrepancy between theoretical and experimental  $\kappa_{ph}$ is not a matter of phonon spectrum, but due to  a scattering mechanism neglected by the theory.

One can extract the temperature dependence of phonon thermal diffusivity by combining the two sets of data and using $D_{ph}=\kappa_{ph}/C_{ph}$. It is shown in Fig\ref{fig2}.c (left y-axis) and reveals three distinct regimes. Above 15 K, $D_{ph}$ decreases with increasing temperature. In the temperature window between 2 K and 10 K, $D_{ph}$ displays a maximum and minimum, both strongly dependent on sample size. Finally, below 1 K, thermal diffusivity becomes sample-independent again and continues its enhancement without saturation down to 0.1 K. 

In the high-temperature regime, phonon-phonon Umklapp collisions set the magnitude of thermal diffusivity. The intermediate temperature regime is analogous  to what has been observed in several other solids, including Bi~\cite{kopylov1973} and black phosphorus~\cite{machida2018}, two other column V elements,  and diagnosed as a signature of phonon hydrodynamics~\cite{beck1974}. In contrast to those materials, the low-temperature thermal diffusivity of Sb does not become not ballistic, but returns to an intrinsic behavior.

\begin{figure*}
\begin{center}
\makebox{\includegraphics[width=1\textwidth]{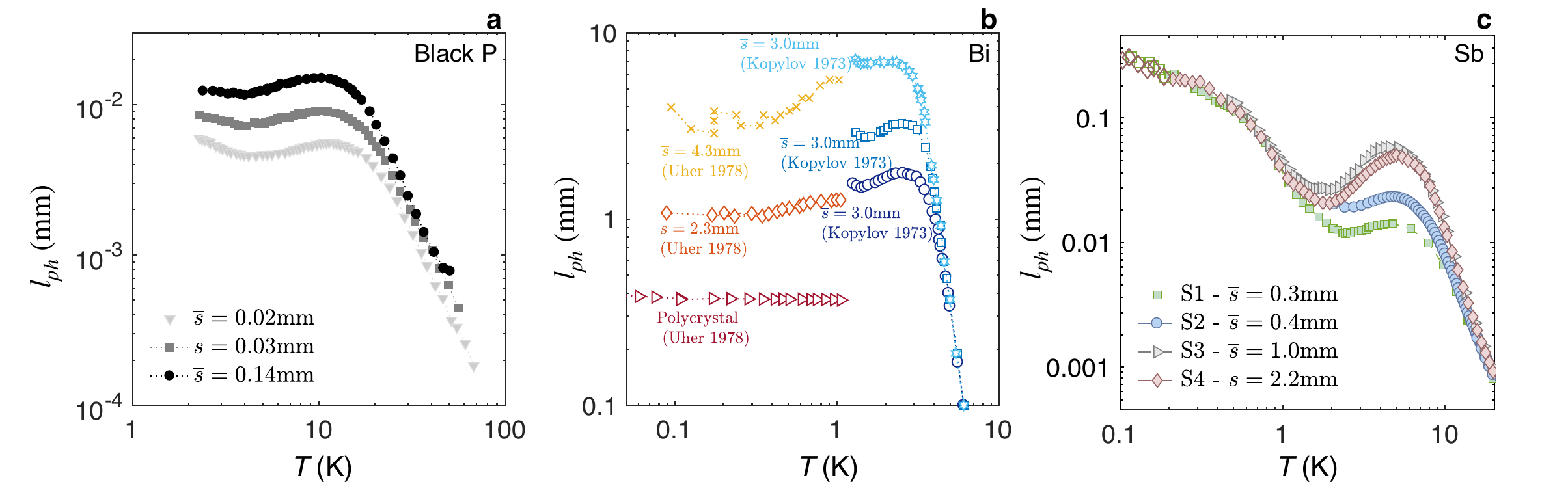}}

\caption{\textbf{The phonon mean-free-path is short and intrinsic} The temperature dependence of the phonon mean-free-path, $\ell_{ph}$ in \textbf{a} black phosphorus \cite{machida2018}, \textbf{b} Bismuth \cite{kopylov1973,pratt1978} and \textbf{c} antimony (this work) as a function of temperature. Only in Sb $\ell_{ph}$ does not evolve with the sample size at low temperature and remains well below the sample thickness.}
\label{Comp_mfp}
\end{center}
\end{figure*}

The phonon mean-free-path can be extracted using the relation $\ell_{ph}=3D_{ph}/{<}v_s{>}$, with ${<}v_s{>}\approx 2900$ m.s$^{-1}$~\cite{epstein1965}. $l_{ph}$ remains well below the typical sample size. Fig.\ref{Comp_mfp} compares the temperature dependence of $\ell_{ph}$ in Sb with  Black phosphorus \cite{machida2018} and Bi \cite{kopylov1973,pratt1978}. Below the hydrodynamic window, phonons in Bi and Black P become ballistic: $\ell_{ph}$ saturates to a value which scales with the crystal size. In the case of Sb, the phononic mean-free-path shows no evidence of ballistic transport down to $T=100$ mK and becomes independent of the sample size.

Thus, phonons in antimony continue to be scattered down to 0.1 K in spite of their long wavelength.  We will show  below that this is due to the coupling between acoustic phonons and electronic quasiparticles. In this new intrinsic regime, a typical phonon goes through numerous momentum-exchanging events with electrons before colliding with the boundary of the crystal.
Note also the large value of the hydrodynamic correction to the mean-free-path which follows from the large ratio of the Poiseuille peak and Knudsen minimum in antimony. It becomes as large as five in the largest samples, compared to 0.4 in the two other solids. Thin graphite \cite{machida2020}, and solid $^4$He \cite{mezhov1966} display a hydrodynamic correction comparable to what was found in samples S3 and S4 of this study.

\begin{figure*}
\begin{centering} 
\makebox{\includegraphics[width=1\textwidth]{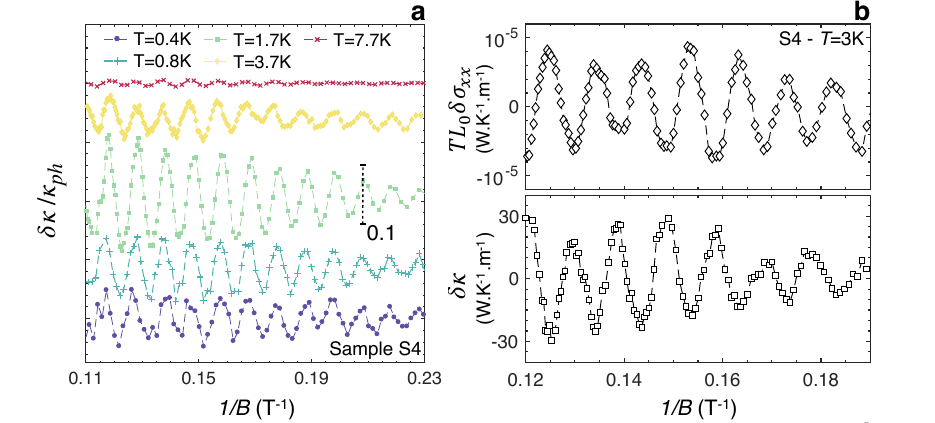}}
    \caption{\textbf{Quantum oscillations of lattice thermal conductivity a} The oscillatory component of the thermal conductivity $\delta\kappa$ is shown (normalized as $\delta\kappa/\kappa_{ph}$) for sample S4 as a function of $1/B$ for various temperatures. Graphs are shifted vertically for clarity. The scale bar corresponds to a relative amplitude of 10\% . \textbf{b} Comparison of the oscillations observed in the thermal conductivity $\delta \kappa$  and in the electrical conductivity (multiplied by the Sommerfeld value and temperature). Note the four orders of difference in amplitude of oscillations in $TL_0 \sigma_{xx}$  and in $\kappa_{ph}$. }
    \label{QO_kappa}
\end{centering}
\end{figure*}

Further evidence for coupling to electrons comes from quantum oscillations of $\kappa_{ph}$ shown in Fig.\ref{QO_kappa}.a. The oscillations, extracted by subtracting a monotonous background, are periodic in $1/B$ and their amplitude depends on the sample size (See the supplement~\cite{SM}). The extracted frequencies~\cite{SM} ($f_{\alpha}=100$T and $f_{\beta}=350-380$T) are in good agreement with previous reports for this configuration~\cite{windmiller1966,brandt1967,fauque2018}. Quantum oscillations of the thermal conductivity have been observed in other semimetals and explained in different manners (See the supplement for details~\cite{SM}). Here, they can safely be attributed to the phononic component. As seen in figure \ref{QO_kappa}.b, their amplitude is four orders of magnitude larger than what is expected for any oscillation in the electronic component, given the amplitude of $\sigma_{xx}$ oscillations and the WF law. In addition, the $\sigma_{xx}$ and $\kappa_{xx}$ oscillations are out of phase. This implies that the enhancement of the electronic density of states (caused by the evacuation of a Landau level) pulls down the phononic thermal conductivity, as would happen if phonons were strongly coupled to electrons.

\begin{figure*}
\begin{centering} 
\makebox{\includegraphics[width=1\textwidth]{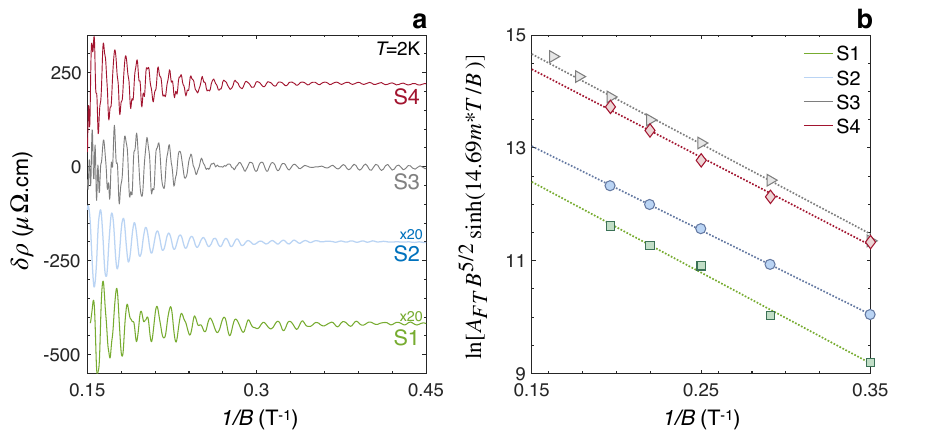}}
    \caption{\textbf{Invariability of Dingle mobility in contrast to transport mobility a} Oscillatory part of the magnetoresistance (for $B$  along trigonal axis) as a function of $1/B$ at $T=2$K in different samples. Curves are shifted vertically and multiplied by a factor $20$ for S1 and S2. \textbf{b)} Dingle analysis of the amplitude $A_{FT}$ of the $100$T-peak of the Fourier transform of $\delta\rho$ in the aforementioned sample. In spite of more than one order of magnitude difference in residual resistivity, the lines are parallel indicating no detectable difference in Dingle mobility. }
    \label{Dingle}
\end{centering}
\end{figure*}

Let us now turn our attention to the electrical properties. Figure \ref{Dingle}.a shows quantum oscillations of the electrical resistivity (the Shubnikov-de Haas effect) in the samples. The Dingle analysis (Figure \ref{Dingle}.b) yields a mobility, $\mu_D$, which is more than two orders of magnitude lower than the transport mobility, $\mu_{tr}$, extracted from the residual resistivity.  A similar observation ( that is $\mu_D \gg \mu_{tr}$)  has been reported in other dilute metals \cite{kumar2017,Liang2015}. In the present case, we find that in all the four samples, in spite of the ten-fold variation in residual resistivity and $\mu_{tr}$,  $\mu_D$ is identical. This indicates that suggesting that $\mu_{tr}$, set by collision time between momentum-relaxing events, varies from sample to sample. On the other hand, $\mu_D$, set by the broadening of the Landau levels, is intrinsically bound. We will see below that this can also fit in our scenario. 

\begin{figure*}
\begin{center}
\makebox{\includegraphics[width=1\textwidth]{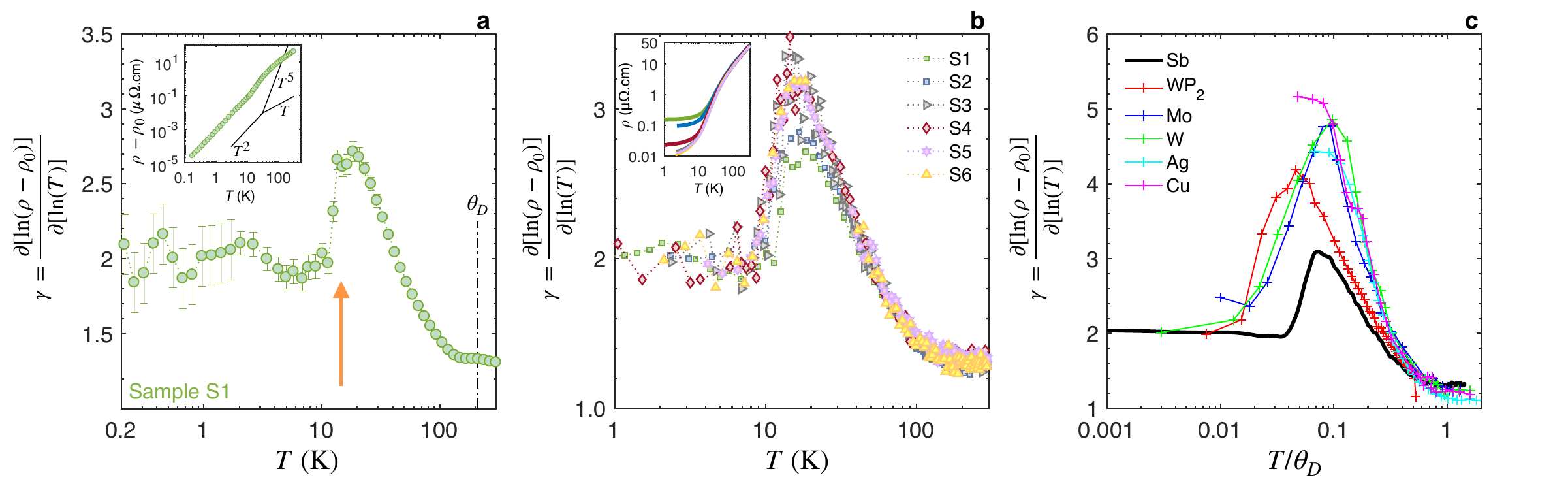}}
    \caption{\textbf{Resistivity exponent a} $\gamma$, the exponent of the inelastic resistivity ($\rho=\rho_0+A_{\gamma}T^{\gamma}$), determined from the logarithmic derivative of $\rho-\rho_0$ in sample S1 as a function of the temperature. Error-bars represent the 95\% confidence bounds. The steady increase in $\gamma$ is abruptly stopped around $T$=15 K, marked by an orange arrow. This is the same temperature at which the phonon mean-free-path becomes size-dependent (arrow in figure \ref{fig2}.c). \textbf{b} The exponent of the inelastic resistivity, $\gamma$,  in six different  Sb samples  as a function of temperature. Inset shows the resistivity of the six samples. \textbf{c} Comparison of the temperature dependence of $\gamma$ in Sb (averaged over the different samples) and five other metals.}
    \label{exponent}
\end{center}
\end{figure*}

The temperature dependence of resistivity provides a crucial piece of information.  Its exponent $\gamma$  ($\rho=\rho_0+A_{\gamma}\times T^{\gamma}$),  does \textit{not} show a $T^5$ behavior  at low temperature (inset of figure \ref{exponent}.a). One can quantify it  by taking the logarithmic derivative  after subtracting residual resistivity:  $\gamma=\frac{\partial\ln(\rho-\rho_0)}{\partial\ln T}$. This procedure was previously applied to extract the exponent of resistivity in cuprates~\cite{Cooper2009}, in heavy fermions \cite{Custers2003} and in strontium titanate \cite{lin2017}.  The temperature dependence of $\gamma$ is shown in Fig.\ref{exponent}.a. In the standard (Bloch-Gr\"uneisen) picture of electron-phonon scattering, resistivity is $T$-linear above the Debye temperature, $\theta_{D}$ (or an effective Debye temperature sometimes called the Bloch-Gr\"uneisen temperature) and evolves towards a $T^5$ behavior upon cooling.  Here,  $\gamma \sim 1$ indeed at high temperature and increases with decreasing temperature. However, this increase is abruptly interrupted around 15 K (orange arrow in Fig.\ref{exponent}.a). Below this temperature, and down to $\approx$0.1 K, $\gamma \sim 2$, with no sign of a higher exponent due to phonon scattering. The same procedure was applied  to six  Sb samples with different residual resistivities. As seen in in Fig.\ref{exponent}.b) all samples show a similar behavior.  This observation confirms the intrinsic nature of  the abrupt shift to $\gamma\sim 2$ and indicates a suppression of resistive electron-phonon scattering at low temperature. The remaining $T^2$-dependent resistivity is associated with electron-electron collisions, which scales with the inverse of the Fermi energy (See the supplement~\cite{SM}).

It is instructive to compare the evolution of the resistivity exponent in Sb with that of other metals. Figure \ref{exponent}.c) shows the temperature dependence of $\gamma$ in Sb (averaged over all six samples samples), and in five other metals. These are semi-metallic WP$_2$ ($n=p=2.5 \times 10^{21}$cm$^{-3}$)~\cite{jaoui2018} , W ($n=p=2 \times 10^{22}$cm$^{-3}$)~\cite{desai1984} , and  Mo($n=p=1 \times 10^{22}$cm$^{-3}$)~\cite{desai1984} as well as noble metals, Cu ($n= 8 \times 10^{22}$cm$^{-3}$) \cite{matula1979} and Ag ($n=6 \times 10^{22}$cm$^{-3}$)~\cite{matula1979}. In almost all cases, when the temperature decreases to one-tenth of the Debye temperature, the exponents increases to 5. This is clearly the case of Ag and Cu. Similar behaviour is found in elemental W and Mo. Nevertheless, because of their lower electronic densities, they display a dominant T-square behavior when $T \ll 0.1\Theta_D$. The prefactor of this T-square resistivity is, however, much smaller than in Sb (See the supplement~\cite{SM}). In contrast to these, the exponent of resistivity in Sb never attains five and suddenly drops to two around 15 K. A similar but less drastic behavior can be seen in WP$_2$. The comparison shows that relaxation of the electron momentum via scattering with phonons vanishes most rapidly in Sb.  This is not surprising given its lower Fermi energy and smaller Fermi surface. 

The  sudden interruption of the growth in the exponent of resistivity in Sb is concomitant with the emergence of the hydrodynamic phonon thermal diffusivity (See orange arrows in Fig.\ref{fig2} and Fig.\ref{exponent}a). There is a simple way of linking the two features. Below 15 K, Umklapp collisions among phonons become rare and  ph-ph collisions conserve momentum. Therefore, a momentum yielded to a phonon during a collision will not be lost to the momentum sink by Umklapp. Rather, it will eventually return to the electron bath through another e-ph collision. The only remaining way for electrons to loose momentum is either elastic collisions or by inelastic resistive collisions with other electrons, still a significant fraction of e-e collisions~\cite{jaoui2021}. As a consequence, $\gamma\approx 2$, from 10 K down to 0.2 K.
This interpretation paves the way to explain other remarkable features found in this study, such as the quantum oscillations of the lattice thermal conductivity and the intrinsic nature of the Dingle mobility, and the fact that the phonon and electron scattering time converge at low temperature (see below). 

\section{Discussion}
\begin{figure*}
\begin{center}
\makebox{\includegraphics[width=1\textwidth]{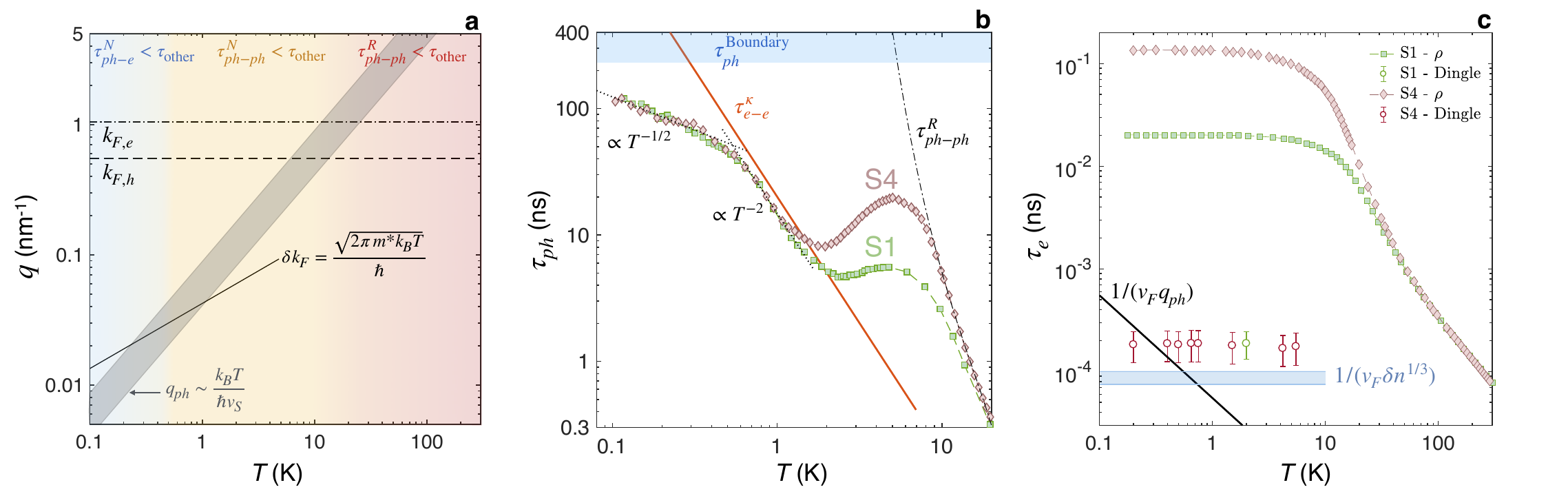}}
    \caption{\textbf{Hydrodynamic regimes. a}  Evolution of the different wavevectors as a function of $T$. Three different regions are identified. \textbf{b} Temperature dependence of the phonon scattering time extracted from the thermal diffusivity of samples S1 and S4 as a function of $T$. Above \textit{T}$\sim$15 K, momentum-relaxing ph-ph collisions dominate. For $1<T<15$ K, momentum-conserving ph-ph collisions lead to a size-dependent decrease in the effective scattering time. Below \textit{T}$\sim$1 K, the most frequent events are momentum-conserving e-ph collisions and the phonon scattering time is close to the e-e scattering time extracted from $\kappa_e$. \textbf{c} Temperature dependence of the electronic scattering time extracted from the electrical resistivity of samples S1 and S4 alongside the Dingle scattering time. Also shown are the typical time between e-ph collisions (solid black line) and the typical time of electron-defect collisions estimated from the Hall resistivity (see supplement for details~\cite{SM}).}
    \label{HD}
\end{center}
\end{figure*}

Usually, phonon hydrodynamics is expected to emerge in a finite temperature window sandwiched between the ballistic and the diffusive regimes~\cite{gurzhi1968,beck1974}. In this hydrodynamic window, the shortest time scale is set by the momentum-conserving collisions between phononic quasiparticles. Our case has a twist. As sketched in figure \ref{HD}.a, below $\sim 15$K, momentum-conserving collisions between phonons become prominent, but as the temperature is reduced further, ph-ph scattering is less frequent than electron-phonon scattering ($\tau^N_{ph-ph} > \tau^N_{ph-e}$). Below $\sim$1 K, the shortest time scale for a phonon corresponds to momentum exchange events with an electron and not with another phonon (or the boundary). In contrast to the common hydrodynamic picture, this new hydrodynamic regime extends down to $\sim$0.1 K, instead of being replaced by a  ballistic regime.

Fig.\ref{HD}.b shows the temperature dependence of the typical phonon scattering time. One can distinguish between two sub-regimes for this e-ph bi-fluid. Above \textit{T}$\approx$0.5 K, the phonon scattering time, $\tau_{ph}$, increases with cooling following ($\tau_{ph}\propto T^{-2}$). More specifically, the extracted $\tau_{ph}$ is quantitively close to   $\tau_{e-e}$, the typical time between two momentum-conserving e-e collisions, extracted from the electronic component of the thermal conductivity (discussed in detail in ref. \cite{jaoui2021}). The most plausible explanation for this is to assume that at least a sizeable fraction of e-e collision events consist in  exchanging phonons. Such an idea has been put forward decades ago in other contexts \cite{Macdonald1980,Gurvitch1980}. Below \textit{T}$\approx$0.5 K, $\tau_{ph}$ saturates towards a $T^{-1/2}$ behavior. This may correspond to a shrinking phase space for an e-ph collision following $\tau^{-1}\propto\frac{q_{ph}}{\delta k_F}\propto T^{1/2}$, where $\delta k_F=\sqrt{2\pi m^* k_B T}/\hbar$ is the thermal thickness of the Fermi surface.

Fig.\ref{HD}.c shows the temperature dependence of the typical electron scattering time, which is much longer than the phonon scattering time. As we saw above, the Dingle analysis of quantum oscillations yields another time scale, which is 2-3 orders of magnitude shorter than the transport time and does not vary from a sample to another. Given that thermal and electric transport by electrons yield a quasi-identical scattering time, the difference with the Dingle time cannot be due to the sampling of small-angle collision events. On the other hand, exchanging back and forth a momentum of $\hbar q_p$ with phonons will widen the Landau levels by an energy of $\sim \hbar v_F q_{ph}$. As seen in Fig.\ref{HD}.c, this gives the right order of magnitude for the experimentally observed Dingle time. However, there is another possible explanation for this. Unavoidable point defects (extrinsic atoms), which give rise to the finite Hall resistivity of this compensated metal, can lead to a short Dingle time (See the supplement~\cite{SM} for details). Introducing controlled disorder will be a way to discriminate between the two possibilities.

In summary, we carried out a detailed study of thermal conductivity in antimony samples of different sizes. We found that the phonon mean-free-path displays a non-monotonous temperature dependence, signaling that phononic heat transport in antimony becomes hydrodynamic  like bismuth and black phosphorus. In contrast to the latter, however, the phonon mean-free-path remains much shorter than the sample size. We argued that this is due to the formation of an electron-phonon bi-fluid, where electrons frequently transmit and receive phonons. Remarkably, around 1 K, the  phonon scattering time and the electron-electron scattering time have the same amplitude and the same temperature dependence.

\section{Acknowledgements}
We thank Yo Machida for discussions. This work was supported by the Agence Nationale de la Recherche (ANR-18-CE92-0020-01 and No. ANR-19-CE30-0014-04), by Jeunes Equipes de l$'$Institut de Physique du Coll\`ege de France and by a grant attributed by the Ile de France regional council. T.L. acknowledges support by the DFG (German Research Foundation) via Project LO 818/6-1. The computational resources were provided by GENCI-CINES (grant 2020-A0090911099) and the Swiss National Superconducting Center (grant s820).

\clearpage
\onecolumngrid
\renewcommand{\thesection}{S\arabic{section}}
\renewcommand{\thetable}{S\arabic{table}}
\renewcommand{\thefigure}{S\arabic{figure}}
\renewcommand{\theequation}{S\arabic{equation}}
\setcounter{section}{0}
\setcounter{figure}{0}
\setcounter{table}{0}
\setcounter{equation}{0}
\newpage
\clearpage

{\large\bf Supplemental Material for ``Formation of an electron-phonon bi-fluid in bulk antimony"}\\

\setcounter{figure}{0}
\section{Samples and methods}
The samples used in this study are presented in table \ref{table1}. They were previously described in detail in ref. \cite{jaoui2021}. The thermal conductivity measurements were performed with home-built one-heater-two-thermometers setups in a Quantum Design PPMS. Different temperature sensors were used for different temperature ranges. Between 80 mK and 4.2 K, Cx-1010 Cernox chips 1010 and RuO$_2$ thermometers were used in a dilution fridge. Between 2K and 40 K, Cx-1030 Cernox chips were used and between 20 K and room temperature, type E thermocouples were used. The overlaps between different sets of experimental data were consistent

Our setups were designed to allow the measurement of both the thermal conductivity, $\kappa$ and the electrical resistivity, $\rho$ with the same electrodes. The thermometers were either directly glued to the samples with Dupont 4922N silver paste or contacts were made using $25\mu$m-diameter silver wires connected to the samples via silver paste (Dupont 4922N). Contact resistance was inferior to $1\Omega$. The thermometers were thermally isolated from the sample holder by manganin wires with a thermal conductance several orders of magnitude lower than that of the Sb samples and silver wires. The samples were connected to a heat sink (made of copper) with Dupont 4922N silver paste on one side and to a RuO$_2$ chip resistor serving as a heater on the other side. Both heat and electrical currents were applied along the bisectrix direction. The heat current resulted of an applied electrical current $I$ from a DC current source (Keithley 6220) to the RuO$_2$ heater. The heating power was determined by $I\times V$ where $V$ is the electric voltage measured across the heater by a digital multimeter (Keithley 2000). The thermal conductivity was checked to be independent of the applied thermal gradient by changing $\Delta T/T$ in the range of 10\%. Special attention was given not to exceed $\Delta T/T\vert_{max}=10\%$. \\The thermometers were calibrated \textit{in-situ} during each experiment and showed no evolution with thermal cycling. Special attention was given to suppress any remanent field applied to the sample and self-heating effects. \\The accuracy of our home-built setups was checked by the recovery of the Wiedemann-Franz law in an Ag wire at $B=0$T and $B=10$T through measurements of the thermal conductivity and electrical resistivity. At both magnetic fields, the WF was recovered at low temperatures with an accuracy of $1\%$ ~\cite{jaoui2018}.

\begin{table*}[h!]
\begin{center}
\begin{tabular}{|c||c|c|c|c|c|c|}
 \hline
Sample & Size (mm$^3$)& RRR  & $\mathrm{\rho_0}$ (n$\Omega$.cm)  & $\mathrm{\overline{s}}$ ($\mathrm{\mu}$m)& $\mathrm{\ell_0}$ ($\mathrm{\mu}$m) & $\rho_0$ $\mathrm{\overline{s}}$ (p$\Omega$ m$^{2}$) \\
\hline\hline
1 & ([0.25$\pm0.05 \times 0.5 \times 4.1$) & 260  & 159 & 350 & 17 & 0.56 \\
\hline
1b & ($0.2\times0.5 \times 4.6$) & 250 & 164 & 320 & 16 & 0.49 \\
 \hline
2 & ($0.4 \times 0.4 \times 4.1$) &  430 & 94.6 & 400 & 28 & 0.38 \\
\hline
3 & ($1.1\times 1.0 \times 10.0$) &  3000 & 13.4  & 1050 & 197 & 0.14  \\
\hline
3$^*$ & ($1.1\times 1.0 \times 7.0$) (cut from 3) & 3000  & 13.4 & 1050 & 197 & 0.14 \\
 \hline
4 & ($1.0\times 5.0 \times 10.0$) &  1700 & 24.1  & 2240 & 110 & 0.54  \\
 \hline
5 & ($3.0\times 1.0 \times 10.0$) & 3700  & 11.1 & 1730 & 238 & 0.19 \\
 \hline
6 & ($1.7\times1.8 \times$ 10.0) & 4200  & 9.8  & 1800 & 270 & 0.18 \\
\hline
\end{tabular}
\caption{\textbf{Details of the samples.} The Sb crystals used in this study were oriented along the bisectrix crystallographic axis. $\overline{s} = \sqrt{\mathrm{width} \times \mathrm{thickness}}$ represents the average diameter of the conducting cross-section. The residual resistivity ratio is defined as $RRR=\frac{\rho_{300K}}{\rho_0}$. The carrier mean free path $\ell_{0}$ was calculated from the residual resistivity and the expression for Drude conductivity assuming three spherical hole and three spherical electron pockets. This is a crude and conservative estimation, because the mean free path of hole-like and electron-like carriers residing in different valleys is likely to differ. Also given is the product of $\rho_0 \overline{s}$, a measure of crystalline perfection.}
\label{table1}
\end{center}
\end{table*}

\section{Separating electronic and lattice components of thermal conductivity}

Our starting point is to assume that  at each temperature the thermal conductivity has two components, The first is not modified by magnetic field, but the second is reduced by magnetic field. Specifically, we found that the field dependence of thermal conductivity at each temperature can be described by the following expression: 
\begin{equation}
\kappa (B)= \kappa_0+\frac{T}{A+CB^n}
\label{separation fit}
\end{equation}
Here, $T$ is temperature and $B$ is the magnetic field. $\kappa_0$, $A$ and $C$ are temperature-dependent parameters and $n\approx 2$ is yet another parameter. As seen in Fig. this expression provides an adequate description of experimental data at all temperatures. It is natural to identify the two components of thermal conductivity as  $\kappa_{ph}=\kappa_0$ and $\kappa_{e}=\frac{T}{A+CB^n}$. Note that the functional form of the latter represents  the expected orbital thermal magnetoresistance of the electronic quasi-particles. 

The separation of the two components leads to Fig.1.c of the main text.  The extracted electronic term combined with electric resistivity quantifies the Lorenz number, which can be compared with the Sommerfeld value. As expected, the WF law is valid at both high temperature and zero temperature limits. 

As shown in the inset of Fig.\ref{fig1}.a The magnetoresistance of antimony is very large at cryogenic temperatures. This is a consequence of the notoriously mobile charge carriers of Sb \cite{fauque2018}. Below 45 K, the thermal conductivity becomes flat above a threshold field. At 5 K and below, a field as low as  1 T enhances the electronic thermal resistivity by many orders of magnitude. Therefore measuring thermal conductivity at 1T yields  lattice thermal conductivity. Subtracting the latter from the zero-field thermal conductivity yields the electronic component.

\begin{figure*}
\begin{center}
\makebox{\includegraphics[width=0.8\textwidth]{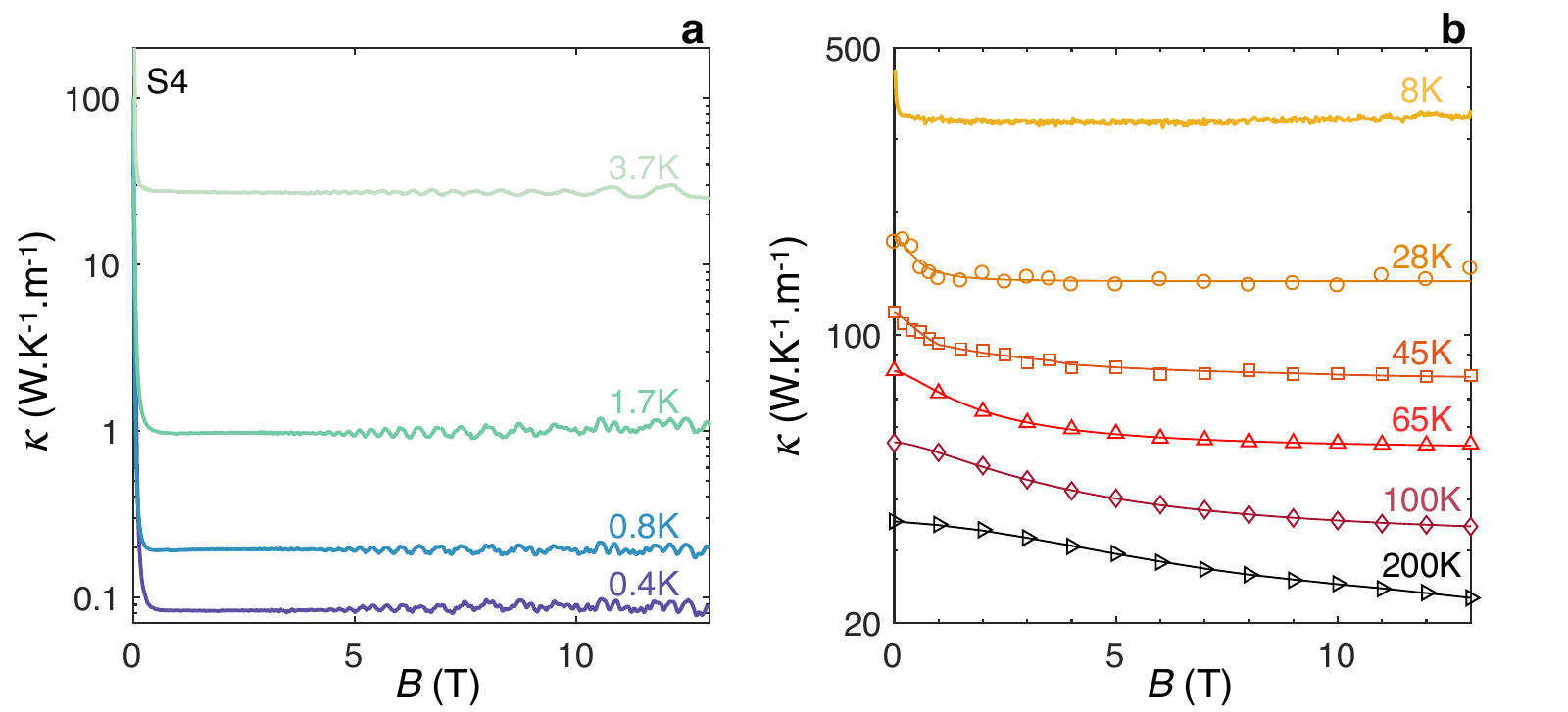}}
    \caption{Evolution of the thermal conductivity of sample S4 with magnetic field for temperatures ranging from 0.4 K to 200 K. The left panel, which focus on temperatures below 5 K, illustrates the suppression of the field-dependent $\kappa$ (i.e. the electronic part) upon application of a fraction of a Tesla to the sample. The remaining constant $\kappa$ (modulo the oscillating part) is associated with the phonon contribution $\kappa_{ph}$. At higher temperatures, featured on the right panel, increasing magnetic fields are required to suppress the electronic component and a clear saturation is not visible above 28 K. The phononic thermal conductivity is then evaluated from a fit to  $\kappa=\kappa_{ph}+T/(\alpha+\beta B^\gamma)$ shown as solid lines.}
    \label{figSM_separation}
\end{center}
\end{figure*}

\begin{figure*}
\begin{center}
\makebox{\includegraphics[width=0.6\textwidth]{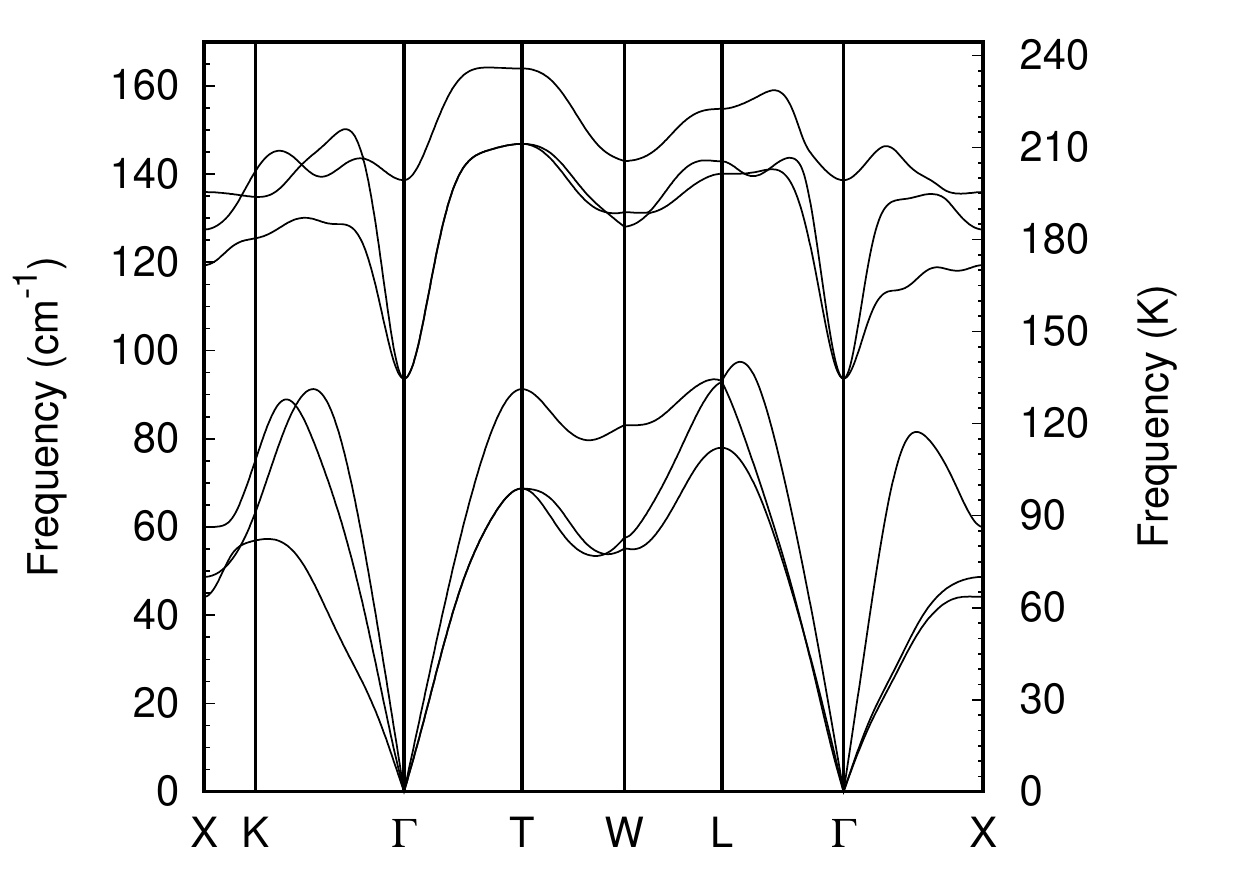}}
\makebox{\includegraphics[width=0.6\textwidth]{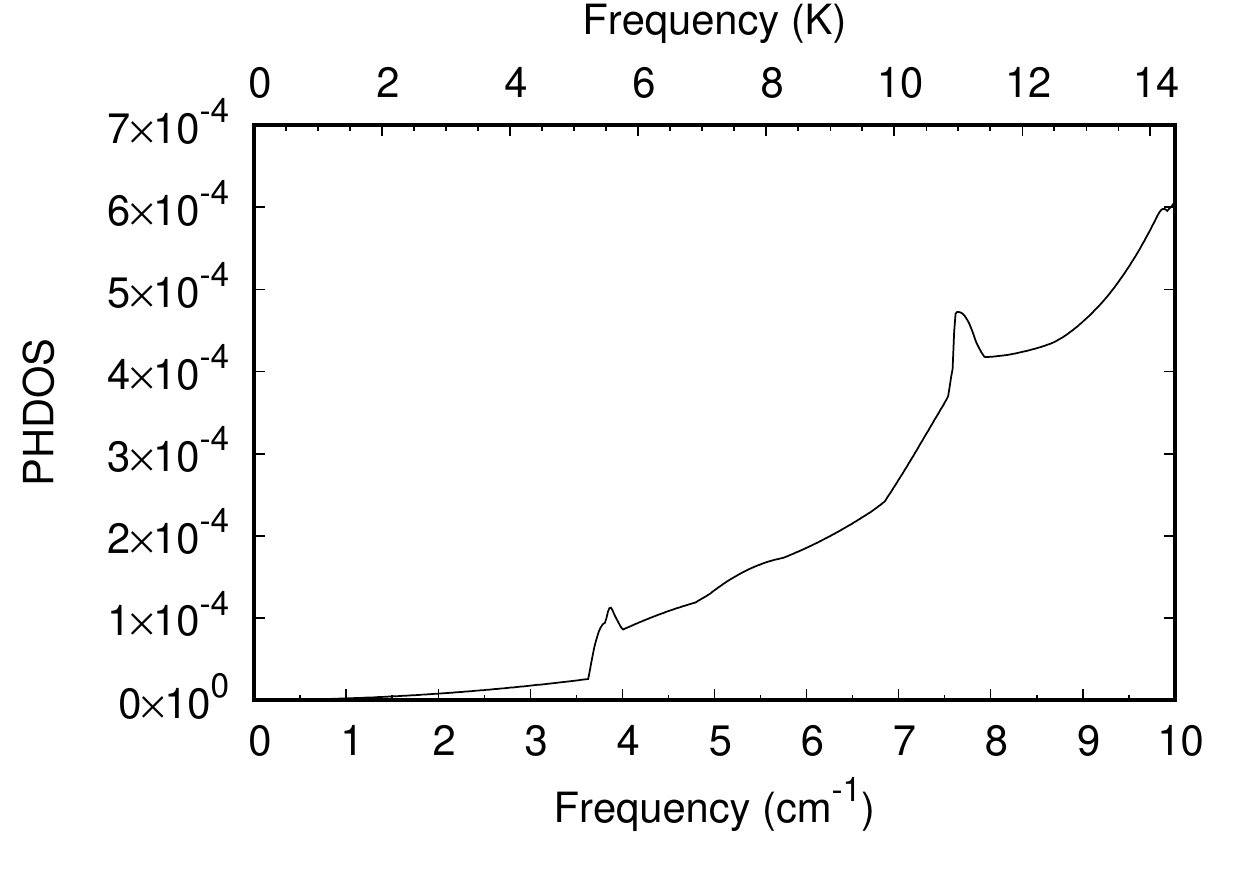}}
    \caption{(Top) Calculated phonon dispersions of Sb plotted along the path $X$ $(0.5, 0, -0.5)$ $\rightarrow$ $K$ $(0.3719, 0, -0.3719)$ $\rightarrow$ $\Gamma$ $(0,0,0)$ $\rightarrow$ $T$ $(0.5,0.5,0.5)$ $\rightarrow$ $W$ $(0.7562, 0.5, 0.2438)$ $\rightarrow$ $L$ $(0.5,0,0)$ $\rightarrow$ $\Gamma$ $(0,0,0)$ $\rightarrow$ $X$ $(-0.5,0,-0.5)$. The coordinates are given in terms of the reciprocal lattice vectors. (Bottom) Calculated phonon DOS of Sb in the low-frequency region. The peaks in the phonon DOS correspond to changes in the slope of the acoustic branches near the Brillouin zone center. This implies that interatomic force constants between atoms at larger distances are sizeable. }
    \label{fig_ph}
\end{center}

\end{figure*}
\section{Theoretical phonon spectrum, specific heat and thermal conductivity}
The phonon calculations were performed using the density functional perturbation theory \cite{dfpt} approach as implemented in the pseudopotential-based planewave code Quantum-{\sc ESPRESSO} \cite{qe}.  We used cutoffs of 75 and 750 Ry for the basis-set and charge-density expansions, respectively.  The exchange and correlation interactions were approximated within the local density approximation. A $20\times20\times20$ $k$-point grid was used in the Brillouin zone integration with a Gaussian smearing of 0.001 Ry.  The dynamical matrices were calculated on a $6\times6\times6$ $q$-point grid, and Fourier interpolation was used to obtain the phonon dispersions and density of states.  We used the fully-relativistic pseudopotential generated by Dal Corso \cite{pslib}, and spin-orbit interaction has been taken into account in our calculations.  The phonon contribution to the specific heat capacity was calculated within the quasiharmonic approximation using the {\sc pyqha} code \cite{pyqha}.  Fully-relaxed structural parameters of $a = 4.4066$ \AA, $\alpha = 58.30^\circ$, and $x = 0.26346$ were used in these calculations. The calculated phonon dispersions and density of states (DOS) are shown in Fig.~\ref{fig_ph}.  

Phonon contribution to the thermal conductivity of antimony was calculated from first principles by taking into account the three-phonon scattering interactions.  We used the frozen-phonon approach as implemented in the {\sc thirdorder.py} \cite{shengbte} code with {\sc quantum-espresso} as the DFT driver to calculate the third-order interatomic force constants (IFCs).  The frozen-phonon calculations were performed on $8\times8\times8$ supercells with a $5\times5\times5$ $k$-point grid.  The calculated second- and third-order IFCs were used to construct the linearized Boltzmann transport equation (BTE)
\begin{eqnarray}
\label{eq:lbte}
\mathbf{F}_{\lambda} & = & \tau_{\lambda} (\mathbf{v}_{\tau} + \mathbf{\Delta}_{\lambda}),
\end{eqnarray}
where $\tau_{\lambda}$ is the relaxation time of the phonon mode $\lambda$ obtained using
perturbation theory, $\mathbf{v}_{\tau}$ is the mode's group velocity, and 
$\mathbf{\Delta}_{\lambda}$ is the correction to the population of the mode $\lambda$ 
obtained from the simple relaxation time approach. Eq.~\ref{eq:lbte} was solved iteratively 
using the {\sc shengbte} package \cite{shengbte} to obtain the lattice thermal conductivity tensor 
$\kappa^{\alpha\beta}$, which is given by the expression
\begin{eqnarray}
    \kappa^{\alpha\beta} = \sum_{\lambda} C_{\lambda} v_{\lambda}^{\alpha} F_{\lambda}^{\beta}.
\end{eqnarray}
Here, $C_{\lambda}$ is the contribution of mode $\lambda$ to the specific heat obtained from the phonon dispersions.  A $28\times28\times28$ grid was used in the solution of the BTE.

Our calculation of thermal conductivity neglects the scattering of phonons with the sample boundary, impurities, and electrons.  A comparison of the calculated and experimental thermal conductivities of antimony shown in Fig.~\ref{fig1}(c) of the main text shows that the calculated values overestimate the experimental ones. The disagreement is small above 100 K, where phonon-phonon scatterings play a dominant role in degrading heat transport.  At lower temperatures, the disagreement becomes larger, suggesting that the neglected scattering mechanisms play a more important role in impeding heat transport.  In insulators such as Si \cite{broido07}, GaAs \cite{lindsay13}, PbTe \cite{shiga12}, SnSe \cite{guo15}, Al$_2$O$_3$ \cite{dongre18}, and In$_2$O$_3$ \cite{subedi21}, theoretical calculations that ignore boundary scattering show remarkable agreement with experiments at low temperatures near the peak in thermal conductivity.  Therefore, the overestimation of the thermal conductivity by our calculations in antimony, which is metallic, suggests that the electron-phonon scatterings play a large role in impeding heat transport at low temperatures in this material.

\section{Oscillations of the lattice thermal conductivity}

\begin{figure*}
\begin{center}
\makebox{\includegraphics[width=0.8\textwidth]{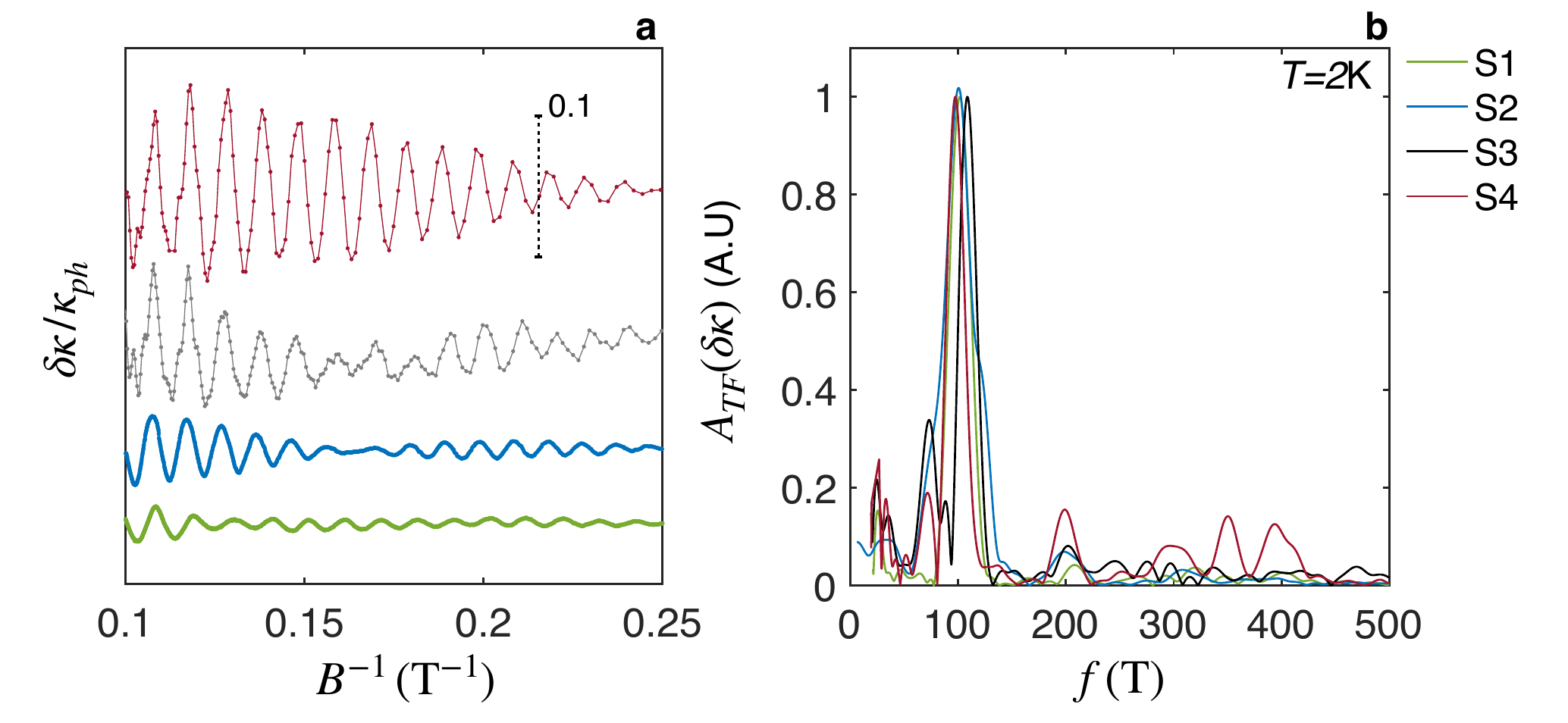}}
    \caption{\textbf{a} The oscillatory part of the thermal conductivity $\delta \kappa$ of the four Sb samples presented in this study is plotted (normalized by $\kappa_{ph}$) as a function of $1/B$ at $T$=2K. The oscillatory part was extracted by removing a polynomial fit to the total $\kappa$. The curves are shifted for clarity. The period of the main oscillation is the same, but a slight misalignment leads to a beating of different frequencies ($80<f<110$T). \textbf{b} Amplitude of the Fourier transform (TF) of $\delta\kappa$ shown in a) for the four Sb samples. The amplitude was normalized by the peak value. The frequency $f=100$T is dominant for all 4 samples.}
    \label{figSM_osc}
\end{center}
\end{figure*}

As discussed in the main text, the lattice thermal conductivity shows quantum oscillations at high magnetic field. Fig.\ref{figSM_osc}.a shows the oscillating component of the thermal conductivity normalized by the phononic thermal conductivity $\kappa_{ph}$, $\delta \kappa/\kappa_{ph}$, in the four different samples studied here at $T=2$K. The amplitude of the oscillations is sample dependent. The amplitude of the oscillation is largest in the larger samples (S3 and S4). The Fourier transform of $\delta\kappa$ (Fig.\ref{figSM_osc}.b) reveals a common frequency $f=100$ T between the different samples. Small variation around this central value are associated with a small misalignment of the magnetic field along the trigonal axis of the device. This frequency is reminiscent of the frequency $f=100$T observed in the oscillations of the magnetoresistance (see \cite{fauque2018} and references within) and associated with the hole pockets of the FS.

Similar oscillations of the thermal conductivity, periodic in inverse of magnetic field, were reported half a century ago, and poorly investigated, in Bi \cite{steele1955}, Sb \cite{long1965} and graphite \cite{ayache1980}. More recently, they have been observed in a variety of semimetals : NbP (and attributed to an ambipolar contribution) \cite{stockert2017}, TaAs (related to variations of the electron-phonon coupling) \cite{xiang2019}, TaAs$_2$ \cite{rao2019}, NbAs$_2$ \cite{rao2019} and in magneto-acoustic properties such as in NbP \cite{schindler2020}. 

\section{Dingle analysis of the quantum oscillations}

\begin{figure*}
\begin{center}
\makebox{\includegraphics[width=0.8\textwidth]{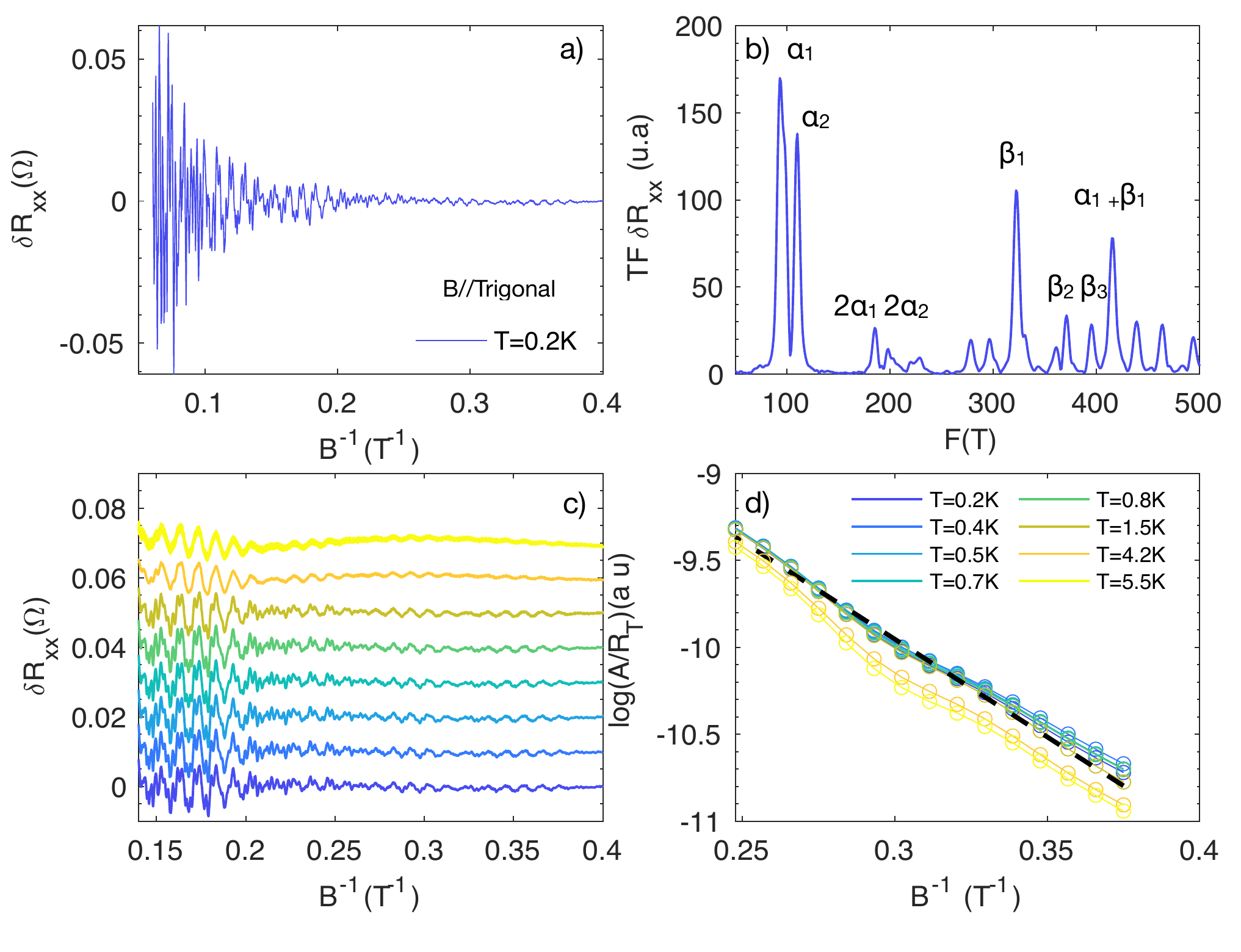}}
    \caption{\textbf{Dingle analysis. a} Trace of the quantum oscillations (QOs) measured in $R_{xx}$ after a polynomial subtraction of the background in S4 at $T=0.2$K as a function of $B^{-1}$. \textbf{b} Fourier transformation of $\delta R_{xx}$ shown in a). \textbf{c} Temperature dependence of the trace of the QOs at low magnetic field : the onset of the lowest frequency ($f_{\alpha}$) is independent of the temperature below 4K. \textbf{d} Dingle plot for the different temperatures sweeps.}
    \label{figSMQO}
\end{center}
\end{figure*}

The Dingle temperature discussed in the main text has been extracted from the analysis of quantum oscillations in resistance ($R_{xx}$). The oscillating component $\delta R_{xx}$ (determined after the subtraction of a background by a polynomial fit) and its Fourier transform (TF) are shown at $T=0.2$ K on Fig.\ref{figSMQO}.a and Fig.\ref{figSMQO}.b respectively. The TF spectrum is dominated by two mains frequencies : $f_{\alpha} = 100$ T and $f_{\beta}\approx350$ T, which correspond respectively to the hole and electrons pockets. Due to a small misalignment of the magnetic field with the trigonal direction both frequencies are split. The sub-frequencies are labelled $\alpha_{i}$ and $\beta_{j}$. A zoom on the low magnetic field oscillations (shown in Fig.\ref{figSMQO}.c) shows that the onset for the emergence of these quantum oscillations ($\mu_{D}B\approx1$ where $\mu_{D}$ is the Dingle mobility) is almost unchanged from $T = 0.2$ K up to $5$ K and that $\mu_{D}$ is of the order 0.4 T$^{-1}$. 

This is confirmed by the Dingle analysis of the oscillations. Fig.\ref{figSMQO}.d shows the so-called "Dingle plot" where $\log(A/R_{T})$, where A is the amplitude of the $F_{\alpha}$ peak in the TF and $R_T = \frac{X}{sinh(X)}$ where $X$ =$\frac{\alpha m^*T}{B}$ with $\alpha=14.694$ and $m^{*}$=0.1m$_0$, is plotted as a function of the average magnetic field of the window on which the TF is done. In this analyse a sliding window of two Tesla has been used. In the context of the Lifshitz-Kosevich formalism \cite{shoenberg2009}, we expect that :  
\begin{equation} \log(\frac{A}{R_T})=-\alpha T_D m^*\times \frac{1}{B}
\label{TD}
\end{equation}

For different temperatures, the field dependence of $\log(\frac{A}{R_T})$ is linear. From Eq.\ref{TD}, we find a Dingle temperature $T_D = 8 \pm 2.5$ K, independent of the temperature and of the residual resistivity. The error bars are estimated including the systematic errors of the background subtraction, the size of the window used for the TF and the fit itself.

\section {Electron-hole compensation and defect concentration}

Due to the charge conservation, a semi-metal is expected to be perfectly compensated with an equal number of electron, $n$, and hole $p$ concentrations. In other words, one expects that in a perfectly stoichiometric sample $n$=$p$. In real materials, however, this charge compensation is inevitably broken by the presence of impurities.

\begin{figure*}
\begin{center}
\makebox{\includegraphics[width=0.8\textwidth]{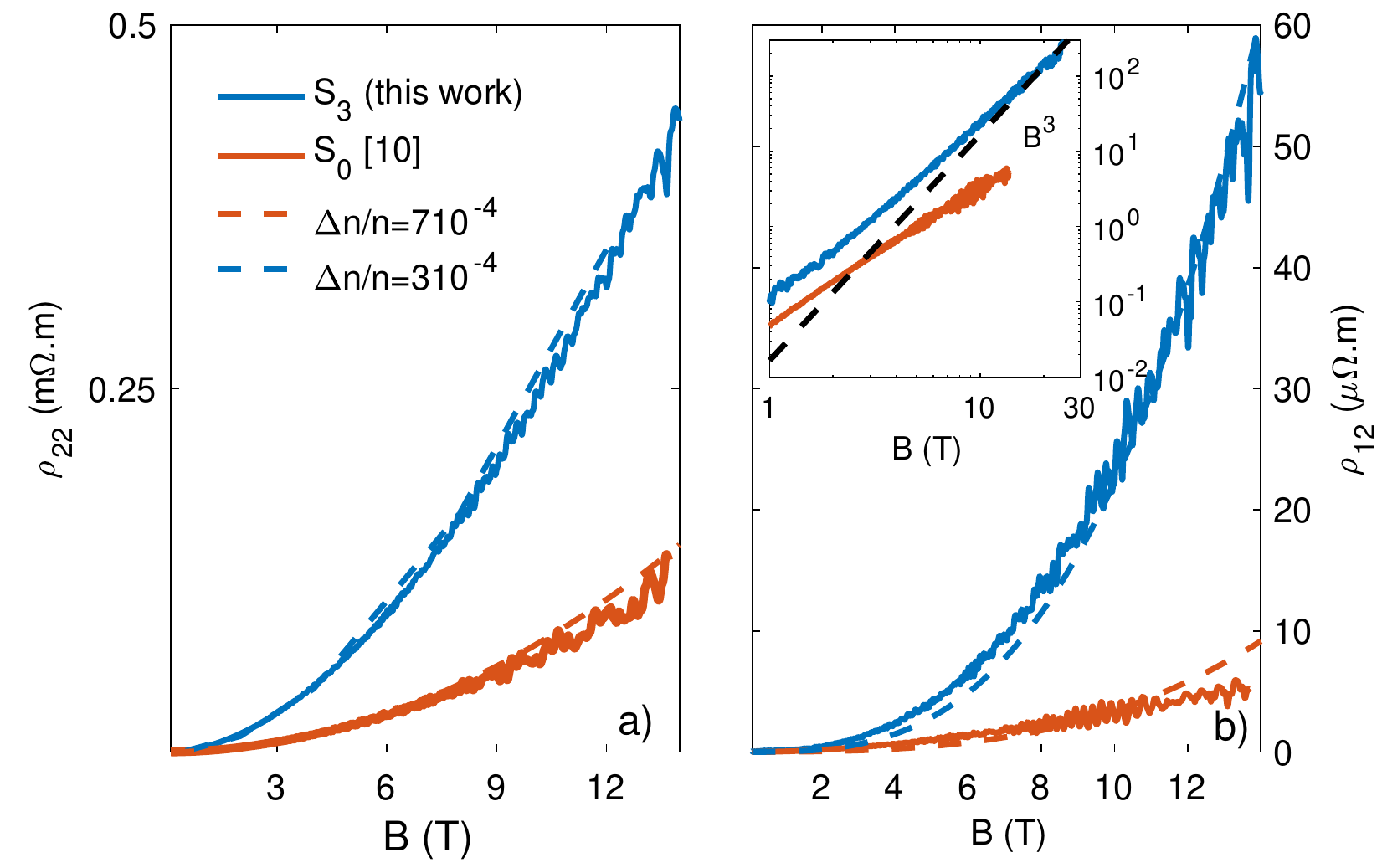}}
    \caption{\textbf{Magnetoresistivity and Hall effect. a} The resistivity ($\rho_{22}$) and \textbf{b} the Hall effect ($\rho_{12}$) for the samples S0 (in red from \cite{fauque2018}) (e) and S3 (in blue) for an electrical current along to the bisectrix direction and a magnetic field parallel to the trigonal direction at $T=2$ K. The dotted lines are fitted using the semi-classical model described in \cite{fauque2018}. Inset in b) : log-log plot of the Hall effect for S0 and S3. The black dotted line indicates a $T^3$ dependence expected in the simplest picture of an uncompensated isotropic electron and hole Fermi surface pockets of same mobility according to Eq.S\ref{BK} when $B<< B_{K}$.}
    \label{figSMHall}
\end{center}
\end{figure*}

A departure from perfect compensation, $\delta n=p-n$ leads to two distinct signatures in the transport properties. The first is found in the magnetoresistance. Assuming perfect compensation, the magnetoresistance is expected to scale as $\propto B^2$ with no saturation. But a finite $\delta n$ would lead to a saturation above a threshold magnetic field of $B_K=\frac{2n}{\delta n \mu}$. This is a simple picture with isotropic pockets of electron and hole of the Fermi surfaces with an identical mobility, $\mu$. The second signature of finite $\delta n$ is a non-linear Hall response in the high-field regime. When $\mu B>>1$, the Hall resistivity, $\rho_{xy}(B)$, can be written as \cite{zhu2015_2} :
\begin{equation}
\rho_{xy}(B)\approx\frac{B}{e} \frac{(n-p)B^2}{\frac{(n+p)^2}{\mu^2}+(n-p)^2B^2}=\frac{B^3}{e(n-p)}\frac{1}{B^2_K+B^2}
\label{BK}
\end{equation}

When the two conditions $B<B_K$ and $\mu B>>1$ are satisfied, the Hall resistivity is expected to show a $B^3$ dependence. A more elaborate model allows to quantify the components of the mobility tensor together with $\delta n$ by a fit to the angle-dependent magnetoresistance \cite{fauque2018}. Fig.\ref{figSMHall} compares the experimental magnetoresistance and  Hall resistivity  in two Sb samples, S0 \cite{fauque2018} ($RRR=400$) and S3 ($RRR=3000$), with a fit using this model. The high-mobility sample (S3) displays a large magnetoresistance and a larger Hall response, indicating that it is more stoichiometric. From these fits, we can estimate $\frac{\delta n}{n}$ to be  $3\times 10^{-4}$ in S3 and $7 \times 10^{-4}$ in S0. Combined with the carrier density of Sb ($n\simeq p\simeq 5.5 \times 10^{19}$ cm$^{-3}$), this implies an impurity density in the range of $1.6-3.8\times 10^{16}$ cm$^{-3}$ and an average distance between impurities of the order of $30-40$ nm. This is two orders of magnitude longer than the average distance between the Sb atoms ($\approx 0.3$ nm) and corresponds to a purity in the range of 1 ppm or less. 

By dividing the distance between the impurities by the average Fermi velocity ($v_F \approx 4 \times 10^5$ m.s$^{-1}$) one finds $\tau_{e-imp.}\approx0.07-0.1$ ps, as shown in Fig.4.d of the manuscript. This time scale is 2-3 orders of magnitude lower than the scattering time yielded by residual electrical (or thermal) resistivity and close to the Dingle scattering time. It is therefore plausible that scattering by point-like impurities decreases the Dingle mobility by widening the width of Landau levels without affecting the flow of momentum or energy and thus leaving the transport mobility unchanged. 

\section{T-square resistivity in semi-metals}

\begin{figure}
\centering
\makebox{\includegraphics[width=0.7\textwidth]
    {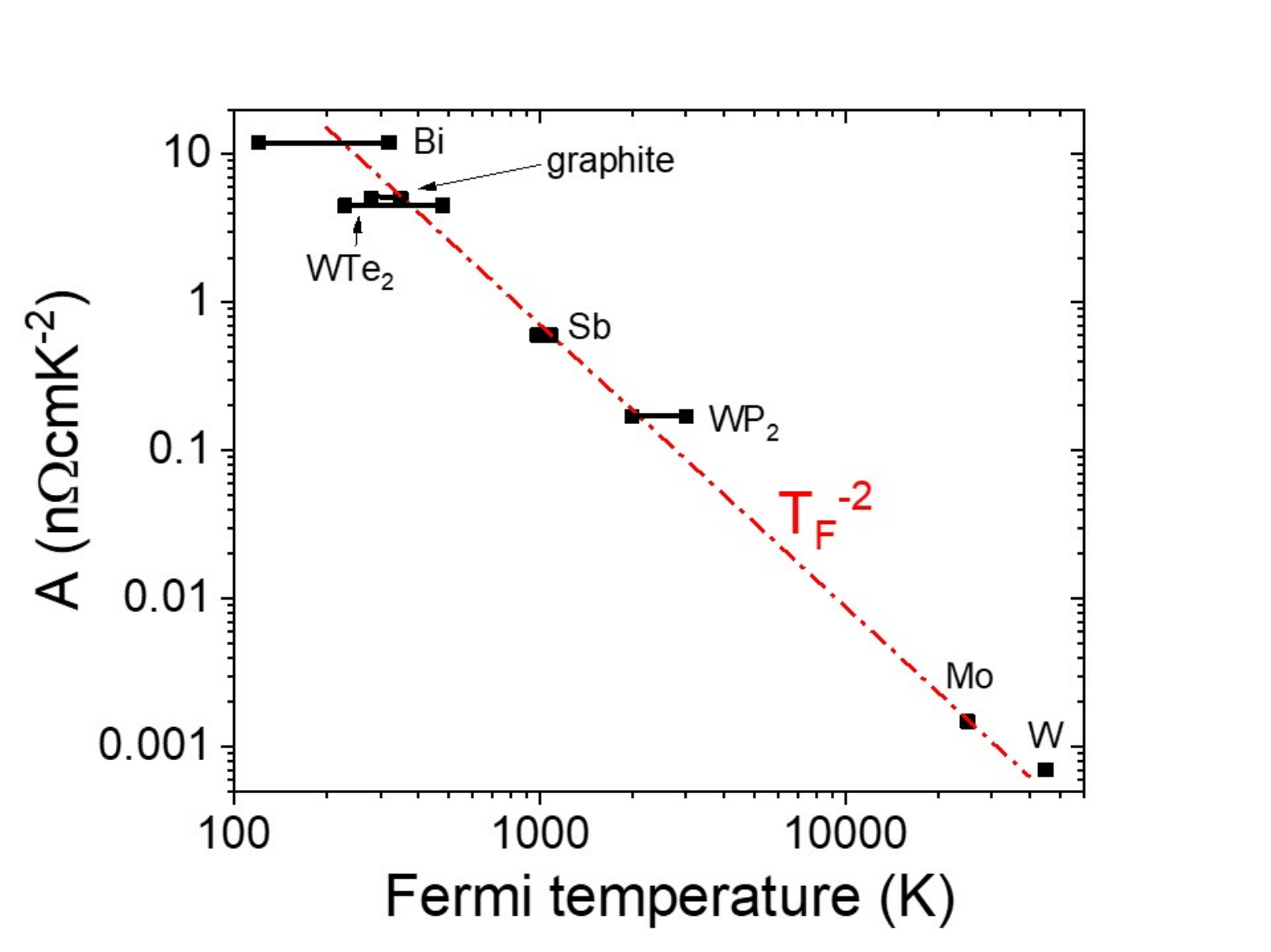}}
    \caption{Prefactor of the $T^2$-resistivity in Bi \cite{uher1977}, graphite \cite{morelli1984}, WTe$_2$ (this work), Sb (\cite{jaoui2021}), WP$_2$ \cite{jaoui2018}, in Mo \cite{desai1984} and in W \cite{wagner1971,desai1984} as a function of the Fermi temperature $T_F$ of electrons and holes. The Fermi energy is as low as 20 meV in Bi and as large as 3 eV in W. Note that carrier density is similar in Sb and WTe$_2$, but not their Fermi energy. This is because electrons and holes are much lighter in Sb.}
    \label{fig_sup_R7}
\end{figure}

Figure \ref{fig_sup_R7} shows the magnitude of the $T^2$-dependent resistivity prefactor $A$ in various semi-metals as a function of their Fermi energy. One can see that the larger the Fermi energy, the larger $A$. This confirms that $A$ scales with the size of the phase space of electron-electron scattering. Indeed, employing twice the Pauli exclusion principle leads to $A\propto(T/T_F)^2$. For a more detailed discussion of this ‘extended Kadowaki-Woods scaling’ see \cite{Wang2020}.

\end{document}